\documentclass[aps,prd,twocolumn,superscriptaddress,showpacs,showkeys,preprintnumbers]{revtex4}

\usepackage[dvips]{graphicx}
\usepackage{epsbox}
\usepackage{amsmath}
\def\slash#1{\not\!#1}

\begin{document}

\preprint{YITP-09-05}
\preprint{RBRC-775}

\title{Spin-3/2 Pentaquark in QCD Sum Rules}


\author{Philipp Gubler}
\affiliation{Department of Physics, H-27, Tokyo Institute of Technology, Meguro, Tokyo 152-8551, Japan}

\author{Daisuke Jido}
\affiliation{Yukawa Institute for Theoretical Physics, Kyoto University, Kyoto 606-8502, Japan}

\author{Toru Kojo}
\affiliation{RBRC, Brookhaven National Laboratory, Upton, New York 11973-5000, USA}

\author{Tetsuo Nishikawa}
\affiliation{Faculty of Health Science, Ryotokuji University, Urayasu, Chiba, 279-8567, Japan}

\author{Makoto Oka}
\affiliation{Department of Physics, H-27, Tokyo Institute of Technology, Meguro, Tokyo 152-8551, Japan}


\date{\today}

\begin{abstract}
The QCD sum rule method is formulated for the strangeness $+1$ 
pentaquark baryon
with isospin $I=0$ and spin-parity $J^\pi = \frac{3}{2}^\pm$. 
The spin-$\frac{3}{2}$ states are considered to be narrower than the spin-$\frac{1}{2}$ ones, 
and thus may provide a natural explanation for the experimentally observed narrow width of $\Theta^+$.
In order to obtain reliable results 
in QCD sum rule calculations, 
we stress the importance of establishing a wide Borel window, 
where convergence of the operator product expansion and 
sufficient low-mass strength of the spectral function are guaranteed. 
To this end, we employ the difference of two independent correlators 
so that the high-energy continuum contribution is suppressed. 
The stability of the physical quantities against the Borel mass is confirmed
within the Borel window.
It is found that the sum rule gives positive evidence 
for the $(I, J^\pi) = (0, \frac{3}{2}^+)$ state with a mass of about $1.4 \pm 0.2$ GeV,
while we cannot extract any evidence for the $(0, \frac{3}{2}^-)$ state.

\end{abstract}


\maketitle

\section{Introduction}
After an earlier prediction by the chiral soliton model \cite{Diakonov}, 
the first positive experimental evidence of $\Theta^+(1540)$ was announced in 2003 by the LEPS collaboration \cite{Nakano1}.
Carrying baryon number $B = +1$ and strangeness $S = +1$, it must be a flavor exotic state and its minimal quark content is $uudd\overline{s}$. 
After the first discovery, numerous theoretical and experimental papers on this pentaquark state have been 
published and the field of hadron spectroscopy has been strongly stimulated by these studies. Nevertheless, the question of the existence of $\Theta^+$ is still a heavily 
disputed issue 
\cite{Barmin1,Barmin2,Staphanyan,McKinnon,Miwa1,Miwa2,Hicks}. 
After reanalyzing the data with higher statistics, the LEPS group has recently again announced the observation of a significant peak \cite{Nakano2}, 
confirming their earlier results, which is a promising sign for further studies of $\Theta^+$.

There have been many attempts to investigate $\Theta^+$ based on approaches closely connected to QCD such as lattice QCD 
\cite{Takahashi,Ishii1,Lasscock1,Ishii2,Lasscock2,Sasaki} and 
QCD sum rules 
\cite{Zhu,Matheus1,Sugiyama,Eidemuller,Ioffe,Kondo,Lee,Nishikawa2,Wei,Matheus2,Kojo1}, but the 
situation does not seem to be completely clear yet, as, for example, no consensus has so far been reached on the spin and parity quantum numbers of $\Theta^+$. 
Explaining the structure of $\Theta^+$ from the first principles of QCD will help to deepen our understanding of hadrons 
in general and particularly of their exotic members.   

An especially peculiar property of $\Theta^+$ is its unnaturally narrow width. Even though its mass lies about $100 \,\mathrm{MeV}$ above 
the $KN$ threshold, the observed width 
appears to be much smaller than for typical baryon resonances. In Ref. \cite{Barmin2} it was reported to be 
even less than $1 \,\mathrm{MeV}$. Several mechanisms have been proposed to explain this narrow width: the strongly correlated diquark model 
\cite{Jaffe,Hosaka}, the $\pi K N$ bound state picture \cite{Bicudo,Kishimoto,Llanes}, the possible isospin $I = 2$ quantum number \cite{Capstick} and the 
possibility of two nearly degenerate pentaquarks \cite{Karliner}. 
Even though all these propositions may be able to explain some properties of $\Theta^+$, they are not yet completely satisfactory. In this paper, we concentrate on another 
explanation, which is the possible spin quantum number $J = \frac{3}{2}$ 
\cite{Takeuchi,Hosaka,Ishii2,Lasscock2,Nishikawa2,Wei}. This spin configuration may provide us with a natural interpretation of the 
narrow width, since in the case of negative parity ($\frac{3}{2}^-$) the only allowed decay is a $KN$ D-wave, which due to the centrifugal barrier 
is strongly suppressed. Additionally, from the 
small wave function overlap further suppression of the decay can be expected. In the case of positive parity ($\frac{3}{2}^+$), the $KN$ decay by the P-wave is allowed and 
the suppression of the width is moderate. Thus other mechanisms for explaining the narrow width may have to be considered.

The main purpose of this work is to investigate $\Theta^+$ states with quantum numbers 
$J^P = \frac{3}{2}^{\pm}$, using the QCD sum rule approach 
\cite{Shifman,Reinders}. There already exist some studies of this problem, in which a similar method was used \cite{Nishikawa2,Wei}, but these works most probably suffer from 
poor convergence of the operator product expansion (OPE) and from the contamination of the continuum contribution in the sum rule. To avoid these problems, we 
employ a method, which 
was first proposed by Kojo, Hayashigaki and Jido \cite{Kojo1} and is especially useful for calculations of exotic states with more than three quarks. In this method, instead of the usual single 
correlator, the difference of two independent correlators is used to construct the sum rule. We find that this procedure 
provides a strong suppression of the continuum contribution of the sum rules. Moreover, to make sure that the OPE converges sufficiently well, we calculate 
the OPE up to dimension 14.

The paper is organized as follows. In section II, we briefly review the basic ideas of the QCD sum rule approach 
and explain the recently introduced improvement of the QCD sum rule method, by which the reliability of our results is increased substantially. In section 
III, the results of the pentaquark mass calculated from the sum rules are presented. We also study the parity of the obtained pentaquark using the parity projected 
sum rule in the chiral limit. 
Finally the conclusion is given in section IV.  

\section{Formulation}
\subsection{QCD sum rules for spin-3/2 particles}
In this section, the basic concepts of the QCD sum rules for spin-3/2 particles are briefly reviewed. Furthermore, 
approximations and conventions used in the calculation are stated.  

QCD sum rules fully exploit the analytic properties of the two point correlation function,
\begin{equation}
\Pi_{\mu\nu}(q) = -i \int d^{4}x e^{iqx} \langle0|T[\eta_{\mu}(x)\overline{\eta}_{\nu}(0)]|0\rangle. \label{eq:cor}
\end{equation}
Here $\eta_{\mu}$ is a Rarita-Schwinger-type interpolating field of a pentaquark, generally carrying components with spin $J = \frac{1}{2}$ and $J = \frac{3}{2}$. 
It is a local operator constructed from the quark degrees of freedom to have the appropriate quantum numbers of the state under investigation.
The imaginary part of $\Pi_{\mu\nu}(q)$ can be expressed as a sum of all hadronic states which couple to the field $\eta_{\mu}$, 
\begin{equation}
\begin{split}
&\mathrm{Im}\Pi_{\mu\nu}(q) \\
&= -\pi \displaystyle \sum_n \delta(q^2-p_n^2)\langle0|\eta_{\mu}(0)|n(p_n)\rangle\langle n(p_n)|\overline{\eta}_{\nu}(0)|0\rangle. \label{eq:imaginary}
\end{split}
\end{equation}
The Lorentz structure of contributions of $J^P = \frac{3}{2}$ states $|n\rangle$ generated by the Rarita-Schwinger field can be obtained as
\begin{equation}
\begin{split}
&\displaystyle \sum_{\mathrm{spin}}
\langle0|\eta_{\mu}(0)|\tfrac{3}{2}^{\pm}(p)\rangle\langle \tfrac{3}{2}^{\pm}(p)|\overline{\eta}_{\nu}(0)|0\rangle \\
&= -|\lambda_{\frac{3}{2}}|^2
\Bigl(g_{\mu\nu} - \frac{1}{3}\gamma_{\mu}\gamma_{\nu} \mp \frac{p_{\mu}\gamma_{\nu}-p_{\nu}\gamma_{\mu}}{3m} -\frac{2p_{\mu}p_{\nu}}{3m^2}\Bigr)
(\slash{p}\pm m),
\end{split}
\end{equation}
where $|\frac{3}{2}^{\pm}(p)\rangle$ denotes a spin-parity $\frac{3}{2}^{\pm}$ hadronic state with mass $m$ and four-momentum $p$, and 
$|\lambda_{\frac{3}{2}}|^2$ is a constant designating the strength of the coupling of $\eta_{\mu}$ to the hadronic state. 
While this expression contains terms proportional to $g_{\mu\nu}$, this is not the case for the states with $J^P = \frac{1}{2}^{\pm}$ (for details see \cite{Hwang}). Therefore 
to project the spin-$\frac{3}{2}$ states out, it is sufficient to just consider the $g_{\mu\nu}$-terms to construct the sum rules,
\begin{equation}
\Pi_{\mu\nu}(q) = g_{\mu\nu}[\Pi_1(q^2)\slash{q} + \Pi_2(q^2)] + \dots \label{eq:evenodd}
\end{equation}
$\Pi_1(q^2)$ (the chiral even part) and $\Pi_2(q^2)$ (the chiral odd part) give two independent sum rules, 
from which we calculate the mass of the state under investigation. As both should in principle lead to the same result, either one 
of them or their combination can be used to carry out the calculation. 

In order to extract physical quantities from the correlation function, we employ the following dispersion relation, which reflects the analyticity of Eq.(\ref{eq:cor}):
\begin{equation}
\Pi_{i}(q^2) = \frac{1}{\pi}\displaystyle \int^{\infty}_{0} ds \frac{\mathrm{Im} \Pi_{i}(s)}{s-q^2}, \label{eq:disp} 
\end{equation}
for $i = 1,2$. 
Possible subtraction terms are neglected here, because they will vanish when the Borel transformation is applied. 
Following the standard technique of QCD sum rules \cite{Shifman,Reinders}, we utilize the ``pole + continuum" ansatz for the imaginary part of the correlator 
in Eqs.(\ref{eq:imaginary}) and 
(\ref{eq:disp}):
\begin{equation}
\mathrm{Im}\Pi_i(s) = \pi|\lambda_i|^2\delta(s - m_{\Theta^+}^2) + \theta(s - s_{th})\mathrm{Im}\Pi_i^{OPE}(s) \label{eq:pole}
\end{equation}
Here $\Pi^{OPE}(s)$ stands for the correlation function calculated with the OPE. It is generally not evident if this ansatz accurately parametrizes the low-energy 
part of the spectral function. Nevertheless, in the case of $\Theta^+$, 
the current positive experimental results indicate that the width of the ground state is very narrow, which allows 
us to express the ground state pole with a $\delta$-function. Furthermore, we can expect that 
because of the suppression due to the centrifugal barrier, the contamination by the $KN$ scattering states is small. 

In order to suppress the higher-order terms of the OPE and the continuum part of the spectral 
function, we make use of the Borel transformation. 
It is defined as
\begin{equation}
L_M[\Pi_i(q^2)] \equiv \lim_{\genfrac{}{}{0pt}{}{-q^2,n \to \infty,}{
-q^2/n=M^2}}
\frac{(-q^2)^{n+1}}{n!} \Bigg(\frac{d}{dq^2}\Bigg)^n 
\Pi_i(q^2), \label{eq:Boreltrans1}
\end{equation}
where $M$ is the Borel mass. 

The result of the OPE for the chiral even part can generally be expressed as
\begin{equation}
\Pi^{OPE}_1(q^2) = \displaystyle \sum^5_{j=0} C_{2j}(q^2)^{5-j}\log(-q^2) + \displaystyle \sum^{\infty}_{j=1} \frac{C_{10+2j}}{(q^2)^j}, \label{eq:ope}
\end{equation}
where the parameters $C_i$ contain vacuum condensates and numerical factors. Substituting Eqs.(\ref{eq:pole}) and (\ref{eq:ope}) 
into Eq.(\ref{eq:disp}), and applying the Borel transformation, we obtain the following expression
\begin{equation}
\begin{split}
&|\lambda_1|^2 e^{-m_{\Theta^+}^2/M^2} \\
&= -\displaystyle \int_0^{s_{th}} ds e^{-s/M^2} \displaystyle \sum^5_{j=0} C_{2j} s^{5-j}  
+ \displaystyle \sum^{\infty}_{j=1} \frac{(-1)^j C_{10+2j}}{\Gamma(j) (M^2)^{j-1}} \\
&\equiv f(M,s_{th}). \label{eq:ope2}
\end{split}
\end{equation}
From the last equation $m_{\Theta^+}$ is then easily obtained:
\begin{equation}
m_{\Theta^+}^2(M, s_{th}) = \frac{1}{f(M,s_{th})} \frac{\partial f(M,s_{th})}{\partial(-1/M^2)}. 
\label{eq:mass}
\end{equation}
In the ideal case, this expression should not depend on the Borel mass $M$, and its dependence on the threshold parameter $s_{th}$ 
should be weak. 

As will be shown later, 
in the actual calculations we apply this method to the difference of two correlators. Therefore, we will not use a single correlator 
as in Eq.(\ref{eq:ope}), but the expression corresponding to Eq.(\ref{eq:Wein}). 
   
\subsection{The importance of the Borel window and its realization}

It is important to assure the reliability of the sum rule by examining the validity of each approximation in the actual calculation. Two critical conditions are 
studied in order.
First, the OPE has to be truncated at a certain order and its convergence is to be checked.
We set the condition so that the contribution of the highest dimensional term is less than 10 $\%$ of all the OPE terms,
\begin{equation}
\frac{L_M\bigl[\Pi^{OPE}_{\text{highest order terms}}(q^2)\bigr]}{L_M\bigl[\Pi^{OPE}_{\text{all terms}}(q^2)\bigr]} \le 0.1. \label{eq:conv}
\end{equation}
The condition is generally satisfied in a restricted region of the Borel mass $M$. As the higher 
dimensional terms get relatively smaller for larger $M$, this condition will set a lower limit for the Borel mass below which the OPE convergence is not 
guaranteed.

The second condition is to suppress the irrelevant high energy contribution above $s_{th}$. We take the condition that the pole contribution is 
dominant ($>50\,\%$) in the sum rule so that unknown contributions from the 
continuum states do not contaminate the result, 
\begin{equation}
\frac{\displaystyle \int^{s_{th}}_{0}ds e^{-\frac{s}{M^2}}\mathrm{Im} \Pi^{OPE}(s)}
{\displaystyle \int_{0}^{\infty}ds e^{-\frac{s}{M^2}}\mathrm{Im} \Pi^{OPE}(s)}\ge 0.5. \label{eq:polecontr}
\end{equation}
This condition tends to be valid generally at a small Borel mass. Thus the condition will set an upper bound for $M$.

The above two conditions often contradict with each other and a valid Borel mass region satisfying both, called a Borel window, may not be obtained. 
In such a case, the sum rule does not give reliable predictions. If the two conditions are satisfied simultaneously 
and thus a valid Borel window is available, the physical quantities can 
be reliably evaluated.

However, in the case of most of the QCD sum rule 
calculations of pentaquark states so far, the two conditions, Eqs.(\ref{eq:conv}) and (\ref{eq:polecontr}) have not been thoroughly checked, and no valid Borel window has been 
established \cite{Matheus2}. Furthermore, 
as has recently been pointed out in \cite{Kojo2}, it is not 
enough just to obtain a stable Borel curve for the physical quantities, as such a stability could be 
produced due to a pseudopeak artifact caused by an inappropriate threshold cut of the spectral function. 
The reason 
for the difficulty of setting up a Borel window 
is first that the convergence of the OPE expansion for a correlator of an interpolating field containing five quarks is considerably slower than in the cases 
of interpolating fields containing only two or three quarks. This makes it necessary to calculate the OPE up to much higher orders 
than in the case of nonexotic hadrons. The second reason 
is the high dimension of the interpolating field of a pentaquark, which causes the continuum part of the spectral 
function to be enhanced. Because of this enhancement, it has been very difficult to obtain a sufficiently high pole contribution.

A solution to this problem was proposed by Kojo \textit{et al.} \cite{Kojo1} in their study of $\Theta^+$ with spin $\frac{1}{2}$. There they made use of the chiral 
properties of two independent interpolating fields and considered, instead of one single correlator, the difference between two correlators of different interpolating 
fields. By this procedure, in analogy to the Weinberg spectral function sum rule \cite{Weinberg}, they realized a strong suppression of the leading orders of the OPE, which mainly contribute to 
the continuum part, and thus obtained a relatively large pole contribution.

We will follow the same lines of reasoning and consider two independent interpolating fields carrying the same quantum numbers \cite{Sasaki},
\begin{equation}
\begin{split}
\eta_{1,\mu}(x) =
&\epsilon_{cfg}[\epsilon_{abc} u^T_a (x)C\gamma_5d_b (x)] \\
&\times [ \epsilon_{def}u^T_d (x)C\gamma_{\mu} \gamma_5d_e (x)]
C\overline{s}^T_g (x),
\end{split}
\end{equation}
\begin{equation}
\begin{split}
\eta_{2,\mu}(x) = 
&\epsilon_{cfg}[\epsilon_{abc}u^T_a (x)Cd_b (x)] \\
&\times [\epsilon_{def}u^T_d (x)C\gamma_{\mu} \gamma_5d_e (x)]
\gamma_5C\overline{s}^T_g (x).
\end{split}
\end{equation}
Here, $a, b,\dots$ are color indices, $C$ is the charge conjugation matrix and T indicates the transposition operation. These fields both 
carry isospin $I = 0$ and have positive intrinsic parity. They are constructed from a scalar diquark, a vector diquark and an anti-strange quark operator in 
the case of $\eta_{1,\mu}$ and from a pseudo-scalar diquark, a vector diquark and an anti-strange quark operator in the case of $\eta_{2,\mu}$, to 
which an additional $\gamma_5$ is added to adjust the parity.
 
A more general operator can be obtained by adopting a linear combination of $\eta_{1,\mu}$ and $\eta_{2,\mu}$:
\begin{equation}
\eta_{\mu}(x)=\cos\theta\eta_{1,\mu}(x)+\sin\theta\eta_{2,\mu}(x).
\end{equation}
Defining the correlator calculated with this general interpolating field as
\begin{equation}
\begin{split}
\Pi(q^2,\theta) =& 
\cos^2\theta\langle\eta_1 \overline{\eta_1} \rangle + \sin\theta\cos\theta[\langle\eta_1 \overline{\eta_2} \rangle+\langle\eta_2 \overline{\eta_1} \rangle] \\
&+\sin^2\theta \langle\eta_2 \overline{\eta_2} \rangle,
\end{split}
\end{equation}
we consider the difference of two 
independent correlators 
\begin{equation}
\begin{split}
\Pi_D(q^2) & \equiv \Pi(q^2,\theta_1)-\Pi(q^2,\theta_2)  \\
&=\sin(\theta_1-\theta_2)\Bigl\{\cos(\theta_1+\theta_2) [\langle\eta_1 \overline{\eta_2} \rangle+\langle\eta_2 \overline{\eta_1} \rangle] \\
&\mspace{18mu}- \sin(\theta_1+\theta_2) [\langle\eta_1 \overline{\eta_1} \rangle-\langle\eta_2 \overline{\eta_2} \rangle] \Bigr\},
\label{eq:Wein}
\end{split}
\end{equation}
and construct the sum rules for this new function $\Pi_D(q^2)$. 
Here, $\langle\eta_i \overline{\eta_j} \rangle$ denotes the relevant part of the correlation function defined in the same way as Eqs.(\ref{eq:cor}) 
and (\ref{eq:evenodd}). 
As seen in Eq.(\ref{eq:Wein}), the sum rule for $\Pi_D(q^2)$ 
depends only on $\theta_1+ \theta_2$, 
because the common 
factor $\sin(\theta_1 - \theta_2)$ drops out in Eq.(\ref{eq:mass}) and therefore does not 
change the value of $m_{\Theta^+}(M, s_{th})$. 
We thus set $\theta_1-\theta_2 = \frac{\pi}{2}$ and $\theta_1+ \theta_2 = \phi$ and subsequently investigate all possible 
values for $\phi$.

It is worth making a few comments on the chiral properties of the interpolating fields. In fact, the sum and the difference of $\eta_{1,\mu}$ and 
$\eta_{2,\mu}$ belong to specific chiral multiplets, as is shown below:
\begin{equation}
\begin{split}
\xi_{1,\mu} &\equiv \eta_{1,\mu} +\eta_{2,\mu} \\
&=  2(u^T_R Cd_R)[(u^T_LC\gamma_{\mu}d_R) - (u^T_RC\gamma_{\mu}d_L)]C\overline{s}^T_R \\
&\mspace{12mu}- 2(u^T_L Cd_L)[(u^T_LC\gamma_{\mu}d_R) - (u^T_RC\gamma_{\mu}d_L)]C\overline{s}^T_L, \\ 
\xi_{2,\mu} &\equiv \eta_{1,\mu} -\eta_{2,\mu} \\
&=  2(u^T_R Cd_R)[(u^T_LC\gamma_{\mu}d_R) - (u^T_RC\gamma_{\mu}d_L)]C\overline{s}^T_L \\
&\mspace{12mu}- 2(u^T_L Cd_L)[(u^T_LC\gamma_{\mu}d_R) - (u^T_RC\gamma_{\mu}d_L)]C\overline{s}^T_R.
\end{split} \label{eq:chiral}
\end{equation}
Here, the color indices have been omitted for simplicity. Eq.(\ref{eq:chiral}) indicates that $\xi_{1,\mu}$ belongs 
to the $(\mathbf{3},\overline{\mathbf{15}}) \oplus (\overline{\mathbf{15}},\mathbf{3})$ multiplet 
with 4(1) right handed and 1(4) left handed quarks, 
and 
$\xi_{2,\mu}$ to the $(\mathbf{8},\mathbf{8})$ multiplet with 3(2) right handed and 2(3) left handed quarks. 
These properties will become important when the difference 
of the correlators is taken.

This can be illustrated by expressing Eq.(\ref{eq:Wein}) in terms of $\xi_{1,\mu}$ and $\xi_{2,\mu}$, which gives
\begin{equation}
\begin{split}
\Pi_D(q^2) 
= &\frac{1}{2}\Bigl\{\cos\phi [\langle\xi_1 \overline{\xi_1} \rangle-\langle\xi_2 \overline{\xi_2} \rangle] \\
&\mspace{18mu}- \sin\phi [\langle\xi_1 \overline{\xi_2} \rangle+\langle\xi_2 \overline{\xi_1} \rangle] \Bigr\}.
\label{eq:Wein2}
\end{split}
\end{equation}
In the second term of this equation both $\langle\xi_1 \overline{\xi_2} \rangle$ and 
$\langle\xi_2 \overline{\xi_1} \rangle$ vanish in the high energy limit, where the chiral 
symmetry is restored. This limit corresponds to the leading perturbative term in the OPE, 
which should therefore similarly cancel. On the other hand, 
it can be understood that the leading orders of the first term proportional to $[\langle\xi_1 \overline{\xi_1} \rangle-\langle\xi_2 \overline{\xi_2} \rangle]$ 
also cancel when an appropriate normalization of $\xi_1$ and $\xi_2$ is chosen. 
That is why we expect the leading orders of 
$\Pi_D(q^2)$ to be suppressed and 
thus to reach a large value for the pole ratio.
As will be shown in the next section, this is in fact the case and we are able to realize a valid Borel window, 
when the OPE is calculated up to sufficiently high orders.
Therefore, 
it is possible to obtain reliable results with the QCD sum rule technique even for a calculation with an interpolating field containing five quarks.

\section{Results}
\subsection{Sum rule for calculating the pentaquark mass} 
We obtain 
the following result for the OPE of the 
chiral even part, in terms of the 
parameters $C_i$ of Eq.(\ref{eq:Wein}). Note, that we here also use $\theta_1-\theta_2 = \frac{\pi}{2}$ and $\phi = \theta_1 + \theta_2$.
\begin{align}
C_0 =& 0, \mspace{18mu} C_4 = \frac{ \langle\frac{\alpha_s}{\pi}G^2\rangle }{2^{16}3^3 5\pi^6}  \cos\phi, \notag\\
C_6 =& \frac{\langle\overline{q}q\rangle^2}{2^83^2\pi^4}\sin\phi 
+ \frac{m_s\langle\overline{s}g\sigma \cdot G s\rangle}{2^{14}3\cdot 5\pi^6} \cos\phi, \notag\\
C_8 =& - \frac{\langle\overline{q}q\rangle\langle\overline{q}g\sigma \cdot G q\rangle}{2^{12}3^3 \pi^4}  (7\cos\phi +172\sin\phi ), \notag\\
C_{10} =& \frac{\langle\overline{q}g\sigma \cdot G q\rangle^2}{2^{14}3^4 \pi^4} (22\cos\phi  +735\sin\phi ) \notag \\
& +\frac{\langle\overline{q}q\rangle^2\langle\frac{\alpha_s}{\pi}G^2\rangle}{2^{10}3^4 \pi^2} (2\cos\phi  -9\sin\phi ) \notag\\
&+ \frac{13m_s\langle\frac{\alpha_s}{\pi}G^2\rangle\langle\overline{s}g\sigma \cdot G s\rangle}{2^{15}3^3 \pi^4}  \cos\phi \notag\\
&+ \frac{m_s\langle\overline{q}q\rangle^2\langle\overline{s}s\rangle}{2^43^2\pi^2} \sin\phi, \notag\\
C_{12} =&-\frac{\langle\overline{q}q\rangle^4}{3^3}\sin\phi \label{eq:even}\\
&-\frac{\langle\overline{q}q\rangle\langle\overline{q}g\sigma \cdot G q\rangle\langle\frac{\alpha_s}{\pi}G^2\rangle}{2^{14}3^4\pi^2} (
65\cos\phi -516\sin\phi ) \notag\\
&+\frac{m_s\langle\overline{q}q\rangle^2\langle\overline{s}g\sigma \cdot G s\rangle}{2^83^3\pi^2} (\cos\phi -30\sin\phi ) \notag \\
&-\frac{7m_s\langle\overline{q}q\rangle \langle\overline{s}s\rangle\langle\overline{q}g\sigma \cdot G q\rangle}{2^73^2\pi^2} \sin\phi,  \notag\\
C_{14}=&-\frac{97\langle\overline{q}q\rangle^3\langle\overline{q}g\sigma \cdot G q\rangle}{2^53^4}\sin\phi \notag\\
&+\frac{m_s\langle\overline{q}q\rangle\langle\overline{q}g\sigma \cdot G q\rangle\langle\overline{s}g\sigma \cdot G s\rangle}{2^{10}3^4\pi^2} (
17\cos\phi -120\sin\phi ) \notag\\
&-\frac{11m_s\langle\overline{s}s\rangle\langle\overline{q}g\sigma \cdot G q\rangle^2}{2^{10}3^3\pi^2}\sin\phi \notag\\
&-\frac{7m_s\langle\overline{q}q\rangle^2\langle\overline{s}s\rangle\langle\frac{\alpha_s}{\pi}G^2\rangle}{2^83^4}\sin\phi. \notag
\end{align}
Here, the definitions $G^2 \equiv G^a_{\mu\nu}G^{a\mu\nu}$ and 
$\sigma \cdot G \equiv \sigma^{\mu\nu} \frac{\lambda^a}{2} G^a_{\mu\nu}$ were used, where $\lambda^a$ are the Gell-Mann 
matrices. $g$ is the coupling constant of QCD, giving $\alpha_s = \frac{g^2}{4\pi}$. The values of the condensates 
and the strange quark mass are given in Table \ref{parameters}.

The coefficient of the leading term, $C_0$, vanishes as we have discussed in the previous section, and the lowest nonvanishing term 
contains a dimension 4 condensate. 
The OPE is 
calculated up to terms with dimension 14 and the conventional vacuum saturation approximation has been assumed. 

We have searched 
all possible values of $\phi$ for a region where a valid Borel window exists and have found such a region around $\phi \sim 0$. 
From Eq.(\ref{eq:Wein2}), it is understood that this region corresponds to the ($\langle\xi_1 \overline{\xi_1} \rangle-\langle\xi_2 \overline{\xi_2} \rangle$)
 component, which interestingly seems to couple strongly to the $\Theta^+$ resonance. 
This also means that the cancellation of the leading term is mainly caused by the appropriate choice of the normalization 
of the operators $\xi_1$ and $\xi_2$, rather than by the restored chiral symmetry. 
To evaluate the final value of $\phi$, and also to obtain the best value of $s_{th}$, 
the following conditions are adopted: 

\begin{description}
\item[1)] A sufficiently wide Borel window exists.
\item[2)] $m_{\Theta^+}(M,s_{th})$ only depends weakly on the Borel mass $M$ and on the threshold parameter $s_{th}$.
\end{description}

The values of $\phi$ and $s_{th}$ that best satisfy 1) and 2) have 
turned out to be $\phi = 0.063$ and $\sqrt{s_{th}} = 2.0\,\mathrm{GeV}$. We will use these values throughout our calculation. 
It should be noted that the threshold parameter $s_{th}$ is chosen to make the sum rule work appropriately, which is not necessarily 
related to the physical continuum threshold or properties of higher resonances.

\begin{table}
\begin{center}
\caption{Values of all the parameters used in the calculation, 
given at a scale of $1\,\mathrm{GeV}$ \cite{Colangelo,Reinders}. 
The parameter $\kappa$ describes the possible breaking of the vacuum saturation approximation and 
is explained at the end of this section.} 
\label{parameters}
\begin{tabular}{lc}
\hline
$\langle\overline{q}q\rangle$ & $-(0.23\pm0.02\,\,\mathrm{GeV})^3$ \\
$\frac{\langle\overline{s}s\rangle}{\langle\overline{q}q\rangle}$ & $0.8\pm0.2$ \\
$\frac{\langle\overline{q}g \sigma\cdot G q\rangle}{\langle\overline{q} q \rangle}$ & $0.8\pm0.1 \,\,\mathrm{GeV}^2$ \\
$\frac{\langle\overline{s}g \sigma\cdot G s\rangle}{\langle\overline{s} s \rangle}$ & $0.8\pm0.1 \,\,\mathrm{GeV}^2$ \\
$\langle\frac{\alpha_s}{\pi}G^2\rangle$ & $0.012\pm0.004\,\,\mathrm{GeV}^4$ \\ 
$m_s$ & $0.12\pm0.06\,\,\mathrm{GeV}$ \\
$\kappa$ & $1\sim2$ \\
\hline
\end{tabular}
\end{center}
\end{table}

\begin{figure}
\includegraphics[width=8cm,clip]{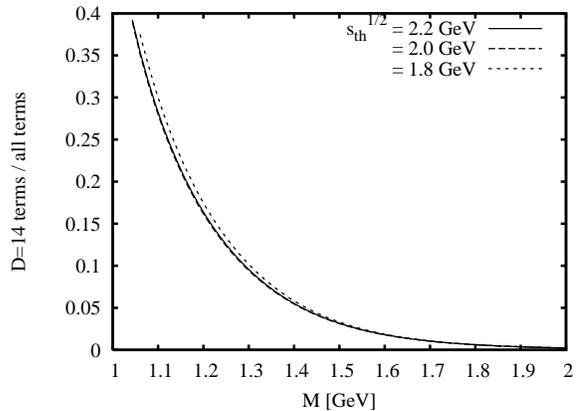}
\caption{The highest term in the OPE divided by the whole OPE, as given in the left hand side of Eq.(\ref{eq:conv}). 
The solid-, dashed- and dotted-curves correspond to $\sqrt{s_{th}} =2.2$, $2.0$ and $1.8\,\mathrm{GeV}$, respectively.}
\label{fig:conv}
\end{figure}

\begin{figure}
\includegraphics[width=8cm,clip]{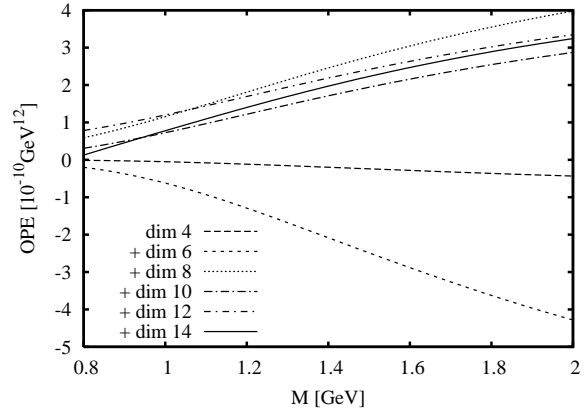}
\caption{Contributions of different dimensions to the right hand side of (\ref{eq:ope2}) for $\sqrt{s_{th}} = 2.0\,\mathrm{GeV}$, added in succession.}
\label{fig:ope}
\end{figure}

First, to demonstrate that the OPE shows a convergent behaviour, the fraction of the highest order terms compared with all the OPE-terms is plotted in Fig. 
\ref{fig:conv} as a function 
of the Borel mass. One sees that the convergence is satisfactory for $M \gtrsim 1.3 \,\mathrm{GeV}$. Additionally, Fig. \ref{fig:ope} shows the contributions 
from the different dimensions to the expression corresponding to the right hand side of Eq.(\ref{eq:ope2}). It is seen that the terms with 
dimension 8 have significant contributions 
to the result. This is an important observation, as the dimension 8 terms have not been included in most previous QCD sum rule calculations of pentaquarks. 

\begin{figure}
\includegraphics[width=8cm,clip]{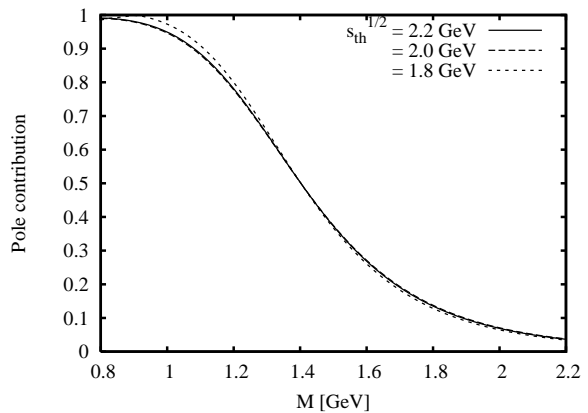}
\caption{The ratio of the pole contribution in comparison with the continuum.}
\label{fig:pole}
\end{figure}

Next, the value of the pole contribution in Eq.(\ref{eq:polecontr}) is shown in Fig. \ref{fig:pole}. As can be read off from the graph the pole contribution is larger than 
50 \% for $M \lesssim 1.4 \,\mathrm{GeV}$. The procedure of taking the difference of two correlators in Eq.(\ref{eq:Wein}) has made it possible to 
obtain such a high value. Altogether, we have achieved a valid Borel window for $1.3 \,\mathrm{GeV}\lesssim M \lesssim 1.4\,\mathrm{GeV}$.
 
\begin{figure}
\includegraphics[width=8cm,clip]{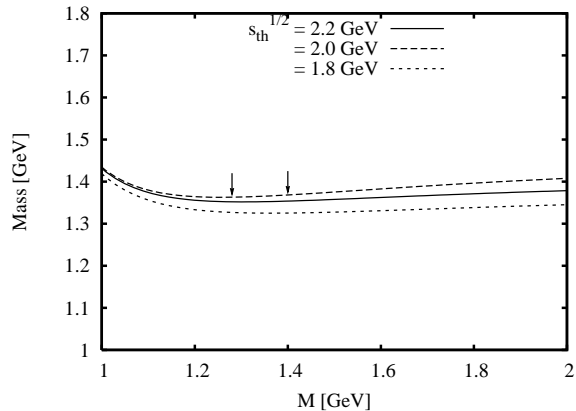}
\caption{The mass of the pentaquark as a function of the Borel mass $M$. The arrows indicate the boundary of the Borel window.}
\label{fig:mass}
\end{figure}

This Borel window is marked by arrows in Fig. \ref{fig:mass}, where $m_{\Theta^+}(M,s_{th})$ is plotted as a function of $M$ for three different threshold parameters. 
One sees that the Borel mass dependence of $m_{\Theta^+}(M,s_{th})$ is weak. 
Moreover, when the threshold parameter $s_{th}$ is varied, this only causes a small shift of the mass, which is much 
smaller than the change of $s_{th}$. These are crucial findings, as they provide convincing support for the ``pole + continuum" 
hypothesis for the spectral function. The result would in fact not depend on $M$ and $s_{th}$ at all if this hypothesis would be completely valid. 

At the same 
time, the results of the last paragraph 
also provide evidence that we are observing an isolated and genuine pentaquark state and not a possible $KN$ scattering state, because 
in the case of a scattering state the dependence of the mass on both $M$ and $s_{th}$ are expected to be stronger. 
The reason for this is the following: if the spectral function contains significant $KN$ scattering states, 
it is natural to expect a rising curve for the mass $m_{\Theta^+}(M,s_{th})$ as the Borel mass increases, because 
the function $m_{\Theta^+}(M,s_{th})$ corresponds to the integrated average of the spectral function from $s=0$ to $s=s_{th}$ with the 
Borel weight $e^{-s/M^2}$. Our result does not show such behaviour. Furthermore, an increase of $s_{th}$ should lead 
to a shift of $m_{\Theta^+}(M,s_{th})$ of similar magnitude if the spectral function contains a large $KN$ background which continues into 
the high energy region. 
Such a shift is not seen in Figs. \ref{fig:mass} or \ref{fig:masschiral}.

A further argument 
in this matter can be made from the dependence of the result on the quark condensate $\langle\overline{q}q\rangle$. The dependence of our result on 
$\langle\overline{q}q\rangle$ is in fact quite small, a change of its value within the error bar just gives a change of the mass value of maximal 
$20\,\mathrm{MeV}$. The mass value actually slightly decreases when the value of $\langle\overline{q}q\rangle$ is increased. 
In the case of a $KN$ scattering state, we would expect an opposite and stronger dependence on $\langle\overline{q}q\rangle$, originating 
from the dependence of the nucleon mass on the quark condensate \cite{Ioffe2}. 
This all suggests that the sum rule is working well, and that the calculated results are reliable.

\begin{figure}
\includegraphics[width=8cm,clip]{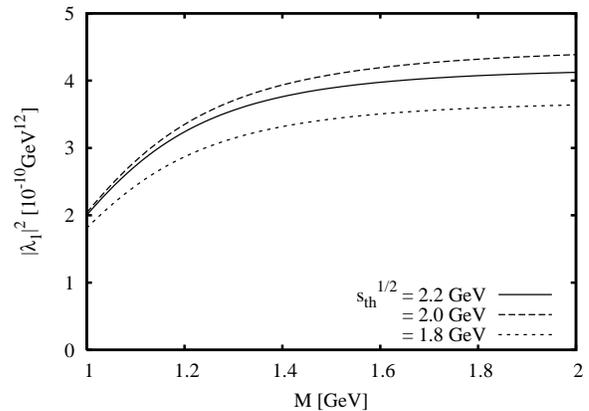}
\caption{The residue $|\lambda_1|^2$ obtained from Eqs.(\ref{eq:ope2}) and (\ref{eq:mass}), for 
different threshold parameters.}
\label{fig:residue}
\end{figure}

As an additional check of the consistency of the sum rules, we calculate the residue $|\lambda_1|^2$, which can be 
obtained from Eqs.(\ref{eq:ope2}) and (\ref{eq:mass}). 
As for the mass, $m_{\Theta^+}$, it should not strongly depend on the Borel mass $M$ or the threshold parameter $s_{th}$.
The result is given in Fig. \ref{fig:residue}. It is seen that 
the stability of $|\lambda_1|^2$ against $M$ is reasonably well and the dependence on $s_{th}$ is moderate.

We finally obtain $m_{\Theta^+} = 1.4 \pm 0.2 \,\mathrm{GeV}$ for the pentaquark mass, which is consistent with the experimental value, although the calculated value is 
somewhat smaller and the theoretical uncertainties expressed as the error bar are large. These uncertainties mainly originate from the possible range 
of the condensates, but also from the breaking parameter $\kappa$ of the vacuum saturation approximation (explained in the next paragraph), 
and finally from the small dependence of the result on $s_{th}$. In fact, while the dependence of the result on the quark and gluon 
condensates is very weak, $m_{\Theta^+}$ significantly depends on the value of the mixed condensate. To be more quantitative, changing the value of  
$\frac{\langle\overline{q}g \sigma\cdot G q\rangle}{\langle\overline{q} q \rangle}$ from $0.8\,\mathrm{GeV}^2$ to $0.9\,\mathrm{GeV}^2$ leads an increase of 
the mass of about $100\,\mathrm{MeV}$.

One last aspect that needs careful consideration is the possible breaking of the vacuum saturation approximation that we have used throughout our 
calculation.
As a test of the validity of this approximation, we have introduced the parameter $\kappa$, which parametrizes the possible violation of 
factorization,
\begin{equation}
\begin{split}
\langle\overline{q}q \overline{q}q\rangle &= \kappa \langle\overline{q}q\rangle^2, \\
\langle\overline{q}q \overline{q}q\overline{q}q\rangle &= \kappa^2 \langle\overline{q}q\rangle^3, \\
\langle\overline{q}q \overline{q}g\sigma \cdot Gq\rangle &= \kappa \langle\overline{q}q\rangle\langle\overline{q}g\sigma \cdot Gq\rangle, \\
\langle\overline{q}g\sigma \cdot Gq \overline{q}g\sigma \cdot Gq\rangle &= \kappa \langle\overline{q}g\sigma \cdot Gq\rangle^2, \\
&\dots
\end{split}
\end{equation}
We have found a mild dependence of the final result on $\kappa$. 
In fact, while varying $\kappa$ in the region of $1\sim2$, the resultant change in the mass of $\Theta^+$ was less than $100\,\mathrm{MeV}$. 
At the same time, the Borel window does not disappear and the conditions for reliable sum rules remain to be satisfied. 
This indicates that the uncertainty introduced by the vacuum saturation approximation is small enough not to change the result qualitatively.

Adding up the main contributions of uncertainty, the dependences on $s_{th}$, 
$\frac{\langle\overline{q}g \sigma\cdot G q\rangle}{\langle\overline{q} q \rangle}$ and $\kappa$, we can conservatively 
estimate the error bar to be $\pm0.2\,\mathrm{GeV}$. 

We have also investigated the $SU(3)_f$ limit ($m_s = 0$, $\langle\overline{q}q\rangle 
= \langle\overline{s}s\rangle$, \dots) to examine the degree of change of the results when this limit is taken.
Our calculations 
show that the results are in fact quite stable and have qualitatively the same behaviour as when the $SU(3)_f$ breaking terms are 
taken into account. The mass $m_{\Theta^+}(M,s_{th})$, for example, decreases only about $50\,\mathrm{MeV}$ in the $SU(3)_f$
limit. This observation will become important in the next section.

\subsection{Determination of the parity}
To determine the parity of the obtained state, we employ the parity-projected sum rule \cite{Jido}. 
The retarded 
Green function is considered in the rest frame:
\begin{equation}
\begin{split}
\Pi_{\mu\nu}^R(q_0) &= -i \int d^{4}x e^{iqx} \langle0|\theta(x_0)\eta_{\mu}(x)\overline{\eta}_{\nu}(0)|0\rangle\Big|_{\vec{q}=0} \\
&\equiv g_{\mu\nu}[\Pi_1^R(q_0)\gamma^0 + \Pi_2^R(q_0)]+ \cdots
\end{split}
\end{equation}
It can be shown that two independent sum rules are derived in a similar way as in the chiral even case. From these sum rules the mass of the ground 
state with positive and negative parity can be obtained:
\begin{equation}
\begin{split}
&|\lambda_{\pm}|^2e^{-(m_{\Theta^+}^{\pm})^2/M^2} \\
&= \frac{1}{\pi} \int_0^{q_0^{th}} dq_0 \bigl[\mathrm{Im}\Pi^R_1(q_0) \pm \mathrm{Im}\Pi^R_2(q_0)\bigr]e^{-q_0^2/M^2}. \label{eq:ope3}
\end{split}
\end{equation}
Therefore, in addition to the chiral even part, the results of the OPE of the chiral odd part are needed here as well. 

However, 
the OPE of $\Pi^R_2(q_0)$ turns out to contain some ambiguous terms proportional to the strange quark mass $m_s$, 
which are attributed to the infrared divergence in the perturbative treatment of $m_s$ \cite{Broadhurst}. 
Here, in order to avoid these ambiguities, 
we 
consider $\Pi^R_2(q_0)$ only in the chiral limit $m_s = 0$. As seen in the last 
section, the results of the calculations of the chiral even part did not qualitatively change when this limit was taken. Therefore 
we can expect the parity-projected sum rule to behave similarly and thus can unambiguously determine the parity of the obtained state. 

The result of the OPE of $\Pi^R_2(q_0)$ is given 
below. Again, we have taken the difference of two correlators with the same conventions of mixing angles as for the chiral even part. 
Here, we use $\phi = 0.063$ as before. 
The parameters $C_i$ are defined similarly as in Eq.(\ref{eq:ope}), replacing $q^2$ by $q_0^2$, and the results are given in the chiral limit.
\begin{align}
C_1 &= 0, \mspace{18mu} C_3 = - \frac{\langle\overline{s}s\rangle}{2^{13}3^3\pi^6}\sin\phi, \notag \\
C_5 &= \frac{5\langle\overline{s}g\sigma \cdot G s\rangle}{2^{13}3^3 \pi^6}\sin\phi, \notag \\
C_7 &= - \frac{7\langle\overline{s}s\rangle \langle\frac{\alpha_s}{\pi}G^2\rangle}{2^{14}3^3 \pi^4} \sin\phi, \notag\\
C_9 &= \frac{5\langle\frac{\alpha_s}{\pi}G^2\rangle\langle\overline{s}g\sigma \cdot G s\rangle}{2^{15}3^4 \pi^4} \sin\phi
 + \frac{\langle\overline{q}q\rangle^2\langle\overline{s}s\rangle}{2^23^3\pi^2}  \sin\phi, \notag \\ 
C_{11} &= \frac{\langle\overline{q}q\rangle^2\langle\overline{s}g\sigma \cdot Gs\rangle}{2^83^3\pi^2} ( 7\cos\phi -18\sin\phi ) \label{eq:odd}  \\
&\mspace{18mu} - \frac{7\langle\overline{q}q\rangle \langle\overline{s}s\rangle\langle\overline{q}g\sigma \cdot Gq\rangle}{2^63^2\pi^2}\sin\phi,  \notag \\
C_{13} &=\frac{7\langle\overline{q}q\rangle^2 \langle\overline{s}s\rangle \langle\frac{\alpha_s}{\pi}G^2\rangle}{2^53^4}\sin\phi \notag\\ 
&\mspace{18mu}-\frac{\langle\overline{q}q\rangle\langle\overline{q}g\sigma \cdot G q\rangle\langle\overline{s}g\sigma \cdot G s\rangle}{2^{10}3^4\pi^2}
(53\cos\phi-168\sin\phi) \notag\\
&\mspace{18mu} +\frac{23\langle\overline{s}s\rangle\langle\overline{q}g\sigma \cdot G q\rangle^2}{2^93^3\pi^2} \sin\phi. \notag
\end{align}
$\Pi^R_1(q_0)$ can be obtained from $\Pi_1(q^2)$ by $\Pi^R_1(q_0) = q_0\Pi_1(q^2_0)$, for details see \cite{Jido}.

\begin{figure}
\includegraphics[width=8cm,clip]{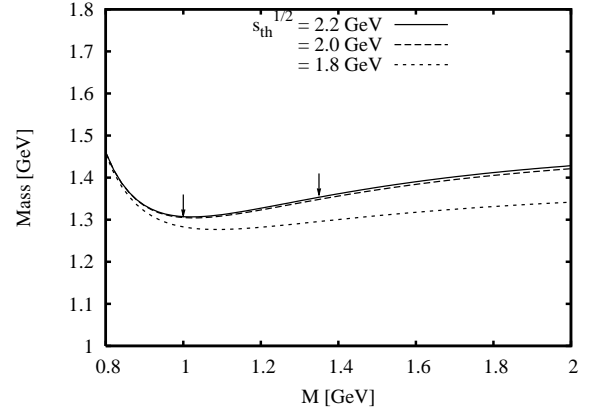}
\caption{The mass of the pentaquark with positive parity as a function of the Borel mass $M$, calculated in the chiral limit. 
The arrows indicate the boundary of the Borel window.}
\label{fig:masschiral}
\end{figure}

In Fig. \ref{fig:masschiral}, we show the plot of the $\frac{3}{2}^+$ pentaquark mass with a valid Borel window 
$1.0 \,\mathrm{GeV} \lesssim M \lesssim 1.3 \,\mathrm{GeV}$.
The obtained mass is consistent but slightly smaller than the one of the chiral even sum rule, probably 
because we take the chiral limit in the parity-projected sum rule. 
In contrast, we can not find any valid Borel window with a stable 
Borel mass curve in the negative parity case. 
We have also checked the relative sign of the residues, independently calculated from the chiral even and chiral odd part, as these residues should 
have the same signs for positive and opposite signs for negative parity states. Both residues turned out to be positive with qualitatively comparable values. These results 
all suggest that the present sum rule predicts a positive parity pentaquark. 

\section{Conclusion}
We have studied the possibility of the quantum numbers of $J^P = \frac{3}{2}^{\pm}$ for the exotic pentaquark $\Theta^+$, using the QCD sum rule method. 
We have obtained a strong suppression of the continuum contribution by constructing the sum 
rules from the difference of two independent correlators. 
Furthermore, we have calculated the OPE up to terms with 
dimension 14, 
which is necessary because of the slow convergence of the OPE of five-quark operators 
and also to increase the components strongly correlated to the ground state. 
We have confirmed previous findings \cite{Matheus2,Kojo1} that 
without these technical improvements, 
studies of exotic states with five 
(or more) quarks may not accomplish meaningful results with the QCD sum rule technique.

One important conclusion of the present work is the confirmation of $IJ^P = 0\frac{3}{2}^{+}$ as possible quantum numbers of $\Theta^+$. 
The numerical result of our calculation is $m_{\Theta^+} = 1.4 \pm 0.2\,\mathrm{GeV}$ for the mass of the pentaquark. 
From the parity-projected sum rule at the chiral limit, we conclude that the parity of the observed state is positive. 
In contrast, we do not find any narrow pole below $2.0\,\mathrm{GeV}$ in the negative parity channel, nor any valid Borel window.

Although the uncertainty of the obtained mass is somewhat large, 
the present result obtained from the QCD sum rules is consistent with the experimental observation by the LEPS group and thus 
suggests that the $\Theta^+$ pentaquark may have spin $\frac{3}{2}$. 
Our conclusions agree with certain earlier quenched lattice results for spin $\frac{3}{2}$ \cite{Lasscock2}. Nevertheless, these results are not confirmed 
by other lattice calculations and no consistent picture has yet emerged. Furthermore, the problem of isolating the pentaquark from the scattering states 
on the lattice 
seems to be a challenging problem. 
From the point of view of explaining the narrow width of $\Theta^+$, our results are not yet conclusive. In other words, 
the mechanisms for explaining the narrow width still have to be clarified and this problem certainly needs further investigation. 

\begin{acknowledgments}
This work was partially supported by KAKENHI, 17070002 (Priority area), 
19540275 and 20028004. 
A part of this work was done in the Yukawa International Project for
Quark-Hadron Sciences (YIPQS). 
P.G. gratefully acknowledges the support from the Ito Foundation of International 
Education Exchange and is thankful for the hospitality of the Yukawa Institute for Theoretical Physics at Kyoto University, 
where part of this work has been completed. 
T.K. is supported by RIKEN, Brookhaven National Laboratory and
the U.S. Department of Energy [Contract No. DE-AC02-98CH10886].
\end{acknowledgments}


\end{document}